\documentclass[prd,amsmath,amssymb,aps,reprint,superscriptaddress]{revtex4-2}
\usepackage{xcolor}
\usepackage[colorlinks=true,linkcolor=blue,urlcolor=purple,citecolor=blue]{hyperref}
\usepackage{graphicx,amsmath}
\usepackage{booktabs}
\usepackage{bm}
\usepackage{float}
\usepackage[normalem]{ulem}
\usepackage{multirow}
\usepackage{ifthen}
\usepackage{times}
\usepackage{orcidlink}

\newcommand{\adsurl}[1]{\href{#1}{ADS}}
\providecommand{\url}[1]{\href{#1}{#1}}

\newcommand{\be}{\begin{equation}}
\newcommand{\ee}{\end{equation}}
\newcommand{\bea}{\begin{eqnarray}}
\newcommand{\eea}{\end{eqnarray}}

\newboolean{editorial}
\setboolean{editorial}{true}
\newcommand{\editorial}[2]{\ifthenelse{\boolean{editorial}}{\textcolor{red}{[\textsf{\textbf{{#1}}}: }\textcolor{blue}{\textsf{{#2}}}\textcolor{red}{]}}{}}

\definecolor{purple1}{HTML}{9654D6}

\definecolor{mangotango}{rgb}{1.0, 0.51, 0.26}

\begin{document}
\title{Quasi-spherical fallback in binary neutron-star mergers: fuel and
  obstacle for extended emission }
%

% ORCID: 0000-0001-7952-7945
\author{Zhiwei Chen \orcidlink{0000-0001-7952-7945}}
% Corresponding author indicated by \email below in REVTeX.
\email{chenzhiwei171@mails.ucas.ac.cn}
\affiliation{Institut f\"ur Theoretische Physik, Goethe Universit\"at, Max-von-Laue-Str. 1, 60438 Frankfurt am Main, Germany}
\affiliation{National Astronomical Observatories, Chinese Academy of Sciences, 20A Datun Road, Beijing 100101, China}

% 0000-0002-0244-2983
\author{Marie Cassing \orcidlink{0000-0002-0244-2983}}
\affiliation{Institut f\"ur Theoretische Physik, Goethe Universit\"at, Max-von-Laue-Str. 1, 60438 Frankfurt am Main, Germany}

% ORCID: 0000-0003-1006-6970
\author{Paramvir Singh \orcidlink{0000-0003-1006-6970}}
\affiliation{Institut f\"ur Theoretische Physik, Goethe Universit\"at, Max-von-Laue-Str. 1, 60438 Frankfurt am Main, Germany}
\affiliation{INAF, Osservatorio di Astrofisica e Scienza dello Spazio,  Via Piero Gobetti 101,  I-40129, Bologna, Italy}

% ORCID: 0000-0002-1330-7103
\author{Luciano Rezzolla \orcidlink{0000-0002-1330-7103}}
\affiliation{Institut f\"ur Theoretische Physik, Goethe Universit\"at, Max-von-Laue-Str. 1, 60438 Frankfurt am Main, Germany}
\affiliation{School of Mathematics, Trinity College, Dublin 2, Ireland}
\affiliation{Frankfurt Institute for Advanced Studies, Ruth-Moufang-Str. 1, 60438 Frankfurt am Main, Germany}

%

%\affiliation{*These authors contributed equally.}

\begin{abstract}

Extended emission (EE) lasting hundreds of seconds has been observed in a
fraction of short gamma-ray bursts containing both nonthermal and thermal
components, and representing a significant puzzle for models involving
binary neutron stars (BNSs). Recent general-relativistic
magnetohydrodynamic {(GRMHD)} simulations of BNS mergers found that the
late-time fallback is approximately quasi-spherical and motivated an
$r$-process powered thermal model for EE. However, the available
radioactive-heating power may be insufficient to sustain the luminosity
inferred by such a model. We therefore propose a fallback-fed
secondary-jet scenario in which a fraction of the fallback material
circularizes on the black-hole equatorial plane and powers the jet, while
polar fallback impedes its propagation. Using a semi-analytic toy model,
we follow the propagation of the jet-head through the fallback envelope
and identify jet-dominated, cocoon-dominated, and choked scenarios. The
boundaries of these regimes depend primarily on the circularization
fraction $f_{\rm circ}$, on the jet half-opening angle $\theta_j$, and on
the polar refilling factor $\xi_{\rm pol}$, with the approximate scaling
$\widetilde{L}\propto f_{\rm circ}/(\xi_{\rm pol}\theta_j^2)$. By
post-processing data from BNS-merger simulations we find that the
fraction of bound fallback material that avoids prompt capture is $f_{\rm
  npc}\simeq0.52$, thus providing an upper bound on the circularizable
fallback supply. Under suitable prescriptions for nonthermal jet emission
and thermal emission from the jet head, the estimated
isotropic-equivalent $0.3$--$10$ keV energies reproduce well both the
observed EE distribution for the jet-dominated branch and its low-energy
end for the cocoon-dominated branch. Overall, these results suggest that
fallback-fed secondary jets represent a dynamically feasible and
energetically consistent explanation of the EE in short gamma-ray bursts.
\end{abstract}
%
% No keywords specified.

\maketitle
\section{Introduction}

Long before the detection of the cornerstone binary neutron star (BNS) merger gravitational wave (GW) event GW170817, \citep{2017ApJ...848L..12A,Kilpatrick2017Sci,2019PhRvX...9a1001A} and its subsequent electromagnetic signal \citep[e.g.,][]{Goldstein2017ApJ,Savchenko2017ApJ,Smartt2017Nature,2017Natur.551...67P,2018ApJ...858L..15D,Hallinan2017Sci,Troja2017Nature}, short gamma-ray bursts (sGRB) had been robustly associated with compact binary mergers involving a neutron star component \citep[e.g.,][]{1994ApJ...431..742D,RevModPhys.76.1143,2014ApJ...788L...8M}. Importantly, while prompt emission from sGRBs typically lasts less than two seconds \citep{kmf+93}, a fraction of
them exhibit a softer extended-emission (EE) component in the X-ray or gamma-ray bands that persists for tens to hundreds of seconds
\citep{2006ApJ...643..266N,2014MNRAS.444L..58V,
2021ApJ...911L..28D,2017ApJ...846..142K}. Additionally,
the spectral properties of EE are diverse:
many events are dominated by nonthermal emission, whereas a significant subset also exhibits a prominent thermal component \citep[e.g.,][]{2020RAA....20..201Z,2024ApJS..275....9C}.

Such long-lasting EE suggests the presence of a central engine or energy reservoir that remains active well beyond the prompt phase. However, both the nature of this engine and the physical mechanism responsible for the EE remain uncertain. One widely discussed possibility is a long-lived magnetar remnant \citep[e.g., ][]{2006Sci...311.1127D,2008MNRAS.385.1455M,2012MNRAS.419.1537B,2024A&A...692A..73G,2026A&A...710A.152P} , whose rotational spin-down or magnetically driven outflow can provide sustained energy injection \citep[e.g.,][]{2015ApJ...805...89L, 2014MNRAS.438..240G, 2016PhRvD..93d4065G,2025MNRAS.544.3883C,Jiang:2025ijp}. A potential challenge for this scenario is to reconcile the long survival of the neutron star with the early production of the powerful ultra-relativistic jet associated with the prompt sGRB. The latter is favored by a sufficiently baryon-clean polar funnel and an ordered large-scale magnetic field, both of which are nontrivial to establish in the immediate post-merger environment
\citep[e.g.,][]{2024ApJ...966L..31M,2020MNRAS.495L..66C,2023PhRvL.131w1402C}. An alternative explanation for the {origin of the} EE has been attributed to the so called ``fallback" emission, where the gravitationally bound material ejected during binary coalescence falls back onto the central remnant and can naturally act as a new source of energy \citep[e.g.,][]{2021ApJ...916L...3M,2021ApJ...922..185I,2022ApJ...936L..10Z,2024MNRAS.528.5952M}.

\begin{figure*}
\centering
\includegraphics[width=1.8\columnwidth, trim=1cm 0cm 0.9cm 0.2cm, clip]{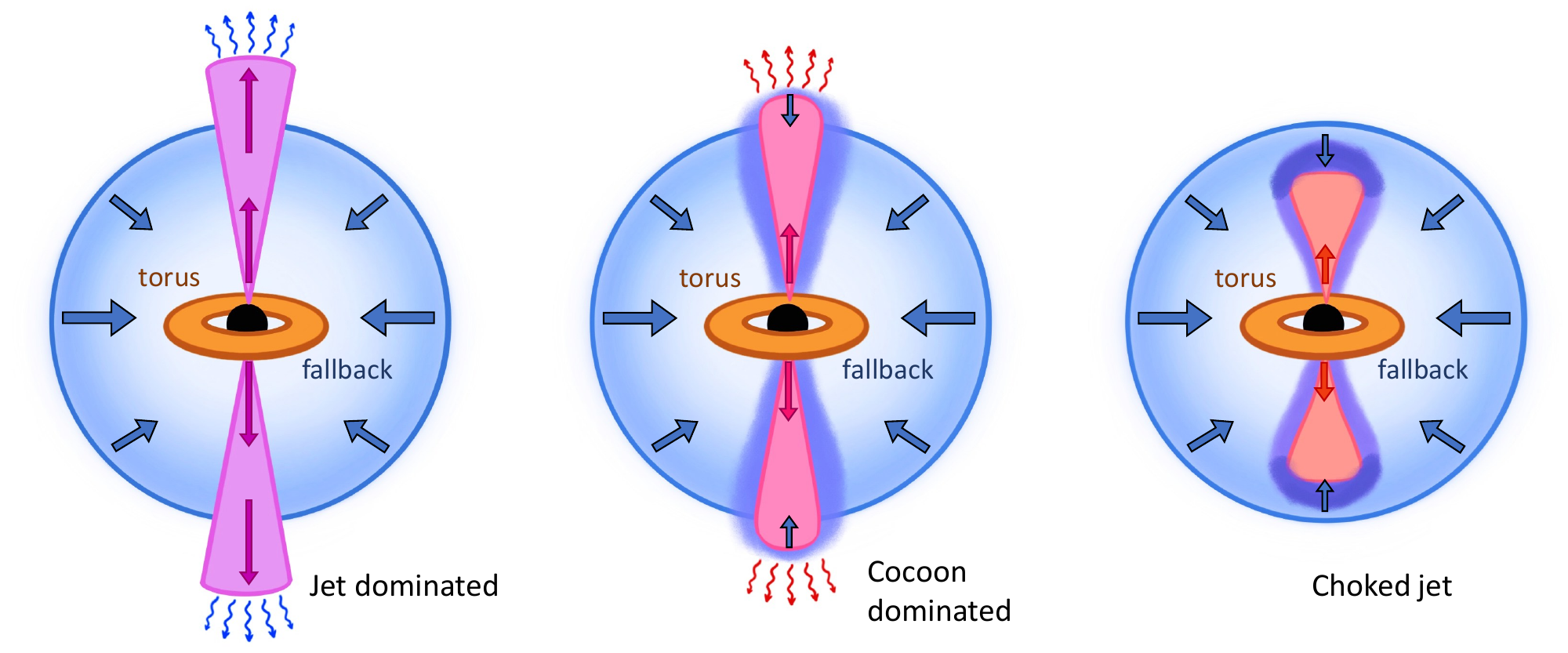}
\caption{Schematic illustration of the fallback-fed secondary-jet model and possible outcome scenarios. The bound merger ejecta continues to fall back toward the central remnant in an approximately quasi-spherical configuration shown in blue. A fraction of this material circularizes and forms an accretion torus, thereby powering a secondary jet, while fallback along the polar direction partially refills the funnel previously cleared by the prompt jet and impedes its propagation. Depending on the competition between the jet power and the density of the polar fallback material, three outcomes are possible: a jet-dominated breakout (left), where the secondary jet successfully penetrates the fallback envelope; a cocoon-dominated breakout (middle), where stronger jet–fallback interaction transfers a substantial fraction of the secondary-jet energy into the surrounding cocoon; and a choked jet (right), where the  jet-tail catches up with the jet-head before a successful breakout is achieved.}
\label{fig:model}
\end{figure*}

Recent general-relativistic magnetohydrodynamic (GRMHD) simulations of BNS mergers \citep[][hereafter MDR24]{2024ApJ...966L..31M} found that a substantial amount of ejecta ($\gtrsim10^{-3}\,M_\odot$, about $\sim 50\%$ of the total ejecta) remains gravitationally bound, while having an approximately quasi-spherical geometry and with a late-time fallback rate $\dot{M}_{\rm FB}\propto t^{-5/3}$. Based on these results, MDR24 proposed that radioactive $r$-process heating of the fallback material could directly power the thermal X-ray emission associated with EE. However, as we show in Appendix~\ref{app:r-process}, their assumption of the local gas temperature as the effective radiative temperature leads to an inferred luminosity that exceeds the available radioactive-heating power. When considering their energy budget and the photon diffusion through the fallback material, the $r$-process heating alone is unlikely to account for the observed EE luminosities. Moreover, a purely thermal fallback-emission model does not naturally explain the substantial population of EEs with predominantly non-thermal spectra \citep[e.g.,][]{2020RAA....20..201Z,2024ApJS..275....9C}. These considerations motivate a different role for the fallback material, {wherein} rather than acting solely as the emitting medium, it may instead serve as the mass reservoir that powers renewed central-engine activity.

One natural realization of such a renewed central-engine activity is a secondary relativistic jet powered by fallback accretion. Fallback-fed secondary-jet models have previously been explored using semi-analytic approaches \citep[e.g.,][]{2022ApJ...936L..10Z,2025A&A...699A.182L}, typically adopting fallback geometries motivated by black-hole--neutron-star (BHNS) mergers. In these geometries, the tidal disruption confines most of the fallback material along the equatorial plane
\citep[e.g.,][]{PhysRevD.92.044028,2021ApJ...915...69D}, while leaving the polar region relatively unimpeded for the secondary jet. However, this geometry does not necessarily apply to BNS mergers, which are expected to constitute a major and likely dominant channel of the observed sGRB population \citep[e.g.,][]{berger+11,nfd+22}. This possibility is further supported by the inferred BNS merger-rate density, which is approximately five times higher than that of BHNS mergers
\citep{ttt+25,2026arXiv260527226T}.
In contrast to the equatorially concentrated fallback expected in BHNS mergers, the GRMHD simulations of MDR24 show that the late-time bound ejecta in BNS mergers becomes approximately quasi-spherical.
This geometry opens a complementary regime in which the same fallback flow can both supply fresh material to the central engine and repopulate the polar region, thereby impeding the propagation of a fallback-fed secondary jet.

In this work, we develop a semi-analytic toy model to investigate the dynamical and energetic viability of a fallback-fed secondary jet as a possible origin of the EE seen in the X-ray observations of sGRBs. We characterize the circularizable fallback supply through the fraction $f_{\rm circ}$ and constrain its plausible range using post-processing of the GRMHD simulations. We then follow the propagation of the secondary jet through the region refilled by polar fallback and identify three possible outcomes: (A) a jet-dominated breakout, (B) a cocoon-dominated breakout, and (C) a choked jet \citep[e.g.,][]{2020A&A...636A.105S,2023MNRAS.520.1111H,2025ApJ...988...30H}. Finally, we show that the isotropic-equivalent X-ray energies of the successful breakout scenario broadly overlap with the observed EE energy range \citep[e.g.,][]{2017ApJ...846..142K}. Our results therefore support fallback-fed secondary jets, regulated by the same quasi-spherical fallback that both powers and obstructs them, as a plausible channel for EE.

This paper is organized as follows. In Section~\ref{sec:model}, we introduce the physical description and formulation of our toy model. In Section~\ref{sec:regimes}, we classify the jet-breakout regimes and estimate their associated X-ray energies. The conclusions and discussions are presented in Section~\ref{sec:cons}.

\section{Model description}\label{sec:model}

The overall physical picture of the proposed toy model is illustrated in Figure~\ref{fig:model}.  The bound merger ejecta continues to fall back toward the central remnant in an approximately quasi-spherical configuration. A fraction $f_{\rm circ}$ of the returning material carries sufficient angular momentum to avoid prompt capture and circularize into an accretion flow, thereby providing the fuel for a secondary jet. Meanwhile, fallback along the polar direction partially refills the channel excavated by the prompt jet and impedes the propagation of the secondary jet. The same fallback material therefore plays two competing roles, simultaneously supplying the central engine and enriching the ambient medium through which the secondary jet must propagate. Depending on the balance between these two effects, the system can result in a jet-dominated breakout, a cocoon-dominated breakout, or a choked jet.

\begin{figure*}
\centering

\includegraphics[width=2.0\columnwidth]{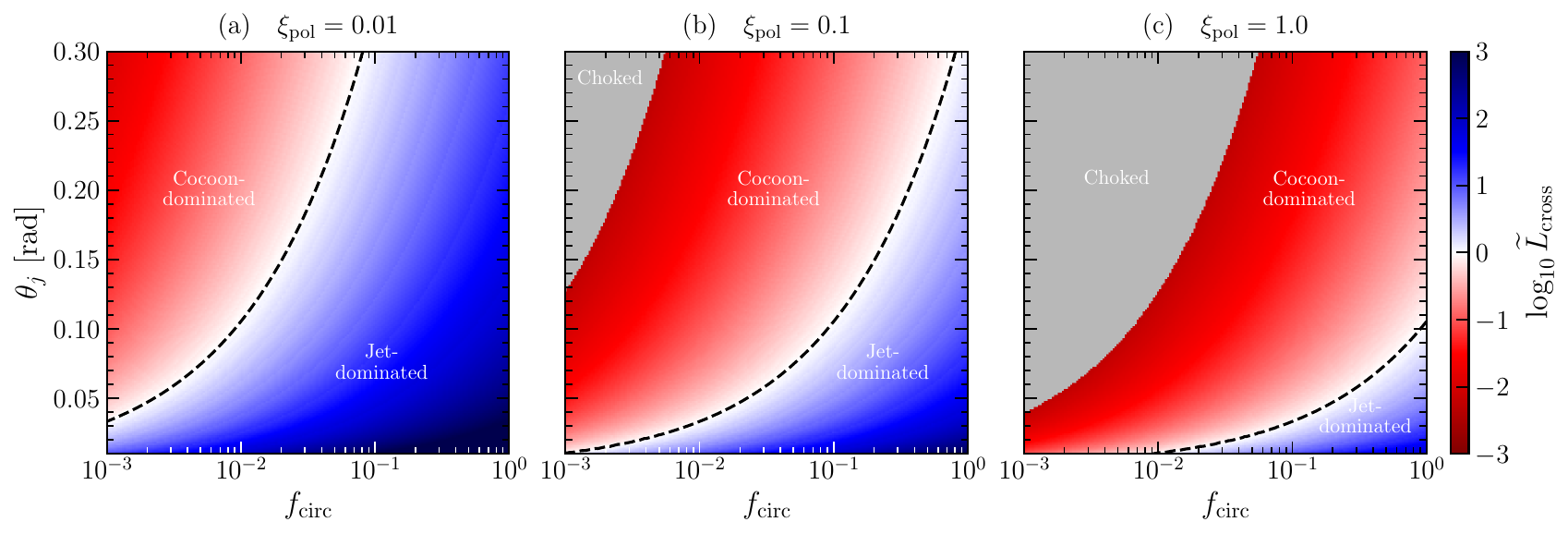}

\caption{Jet-breakout phase diagrams in the $(f_{\rm circ},\theta_j)$ plane for three polar refilling factors: (a) $\xi_{\rm pol}=0.01$, (b) $\xi_{\rm pol}=0.1$, and (c) $\xi_{\rm pol}=1.0$. The color scale shows the dimensionless jet-head luminosity $\log_{10}\widetilde{L}_{\rm cross}$ evaluated at the crossing time and the black dashed line shows the curve for $\widetilde{L}_{\rm cross}=1$. The gray region denotes choked jets with $t_{\rm jet,cross}\leq0$. All other model parameters are fixed at the values listed in Table~\ref{table:parameters}. 
\label{fig:phase_diagram}}
\end{figure*}

In this model, the same fallback flow determines both the power of the secondary jet and the medium through which it propagates. As discussed in MDR24, the fallback rate can be described by the canonical power law
\begin{equation}
    \dot{M}_{\rm FB}(t)=\dot{M}_0 t^{-\alpha},
    \qquad \alpha=\frac{5}{3},
\end{equation}
where $t$ is the time since the merger. Here we normalize $\dot{M}_0$ by the total fallback mass $M_{\rm FB}$ from $t_{\min}=0.1\rm$ s to infinity according to $\dot{M}_0=(\alpha-1)M_{\rm FB}t_{\min}^{\alpha-1}$. Under the ballistic approximation, the leading edge of the fallback envelope evolves as
\begin{equation}
    R_{\rm edge}(t)=R_d
    \left(1+\frac{3}{2}\frac{\sqrt{2GM_c}\,t}{R_d^{3/2}}\right)^{2/3},
\end{equation}
where $M_c$ is the mass of the central remnant and $R_d$ is the reference radius at which the fallback rate is measured. The corresponding density profile, $\rho_{\rm FB}(r,t)$, is reconstructed by applying mass conservation to the free-falling trajectories, as detailed in Section 3.1 of MDR24. For illustration purpose of this work, we adopt the values of $M_{\rm FB}=1.6\times 10^{-3} M_{\odot}$, $M_{c}=2.63M_{\odot}$, and $R_d=300M_{\odot}$ according to MDR24. 

We characterize the two roles of the fallback material through
\begin{equation}
    \dot{M}_{\rm circ}
    =f_{\rm circ}\dot{M}_{\rm FB},
    \qquad
    \rho_{\rm pol}(r,t)
    =\xi_{\rm pol}\rho_{\rm FB}(r,t),
\end{equation}
On the one hand, the circularization fraction $f_{\rm circ}$ measures the global fraction of the fallback material with sufficient angular momentum that can avoid prompt capture and circularize to form an accretion flow. Here we retain $f_{\rm circ}$ as the primary free parameter and explore its effect on jet breakout.  Besides, our GRMHD-based post-processing shows that a fraction $f_{\rm npc}\simeq0.52$ of the bound fallback material is not promptly captured, thus placing an upper bound on $f_{\rm circ}$. Here we consider $f_{\rm circ}=0.052$, $0.26$, and $0.52$ as low, intermediate, and maximal circularization cases, respectively. Notably, the fallback material that is not promptly captured is predominantly distributed outside the bipolar jet funnels, with more than $\gtrsim99.3\%$ lying beyond the polar region for $\theta_j\leq0.3~{\rm rad}$, suggesting that the circularizable reservoir is primarily supplied by nonpolar fallback. More details are provided in the Appendix~\ref{app:fcirc}.

On the other hand, the polar refilling factor $\xi_{\rm pol}$ measures the recovery of the fallback density inside the channel excavated by the first jet. The first jet can shock, accelerate, or unbind the weakly bound polar material and thereby remove or delay material that would otherwise return before the secondary jet is launched. The limit $\xi_{\rm pol}=1$ corresponds to a complete replenishment of the unperturbed quasi-spherical fallback density, whereas $\xi_{\rm pol}\ll1$ describes a strongly depleted polar channel that has only been partially repopulated. Since the interaction between the first jet and the subsequent fallback cannot be followed by the current simulation, $\xi_{\rm pol}$ cannot yet be determined self-consistently. We therefore consider $\xi_{\rm pol}=1$, $0.1$, and $0.01$ as representative cases spanning fully replenished, partially refilled, and strongly depleted polar environments, respectively.

We consider and model the secondary episode of the central engine to be operational over a finite interval, beginning at $t=t_{j,2}$ and lasting for $t_{\rm eng}$. Once this engine becomes active, we assume that its power promptly tracks the circularized fallback supply such that
\begin{equation}
L_j(t)=\eta_j\dot{M}_{\rm circ}(t)c^2
     =\eta_j f_{\rm circ}\dot{M}_{\rm FB}(t)c^2 .
\end{equation}
Here $L_j$ denotes the power of one jet and we adopt a canonical conversion efficiency of $\eta_j=0.1$ \citep{2022ApJ...936L..10Z}. Since the formation of an observable relativistic jet may be delayed relative to the onset of fallback
by disk and magnetic-field evolution, we treat $t_{j,2}$ phenomenologically
\citep[e.g.,][]{2022PhRvD.105j4028S,2024PhRvD.110h3014A}.
We adopt the fiducial values $t_{j,2}=10\ {\rm s}$ and $t_{\rm eng}=100\ {\rm s}$, motivated by the characteristic onset and duration of EE
\citep{2017ApJ...846..142K}. The energy supplied by the engine during this interval is \begin{equation}
    E_j=\int_{t_{j,2}}^{t_{j,2}+t_{\rm eng}}
    L_j(t)\,dt,
\end{equation}
which sets the overall energy budget of the secondary jet. However, whether this energy can power the observable EE also depends on the ability of the jet to successfully propagate through the refilled polar fallback and break out of the envelope \citep[e.g.,][]{2018PTEP.2018d3E02I,2020MNRAS.491.3192H,
2020A&A...636A.105S,2023MNRAS.520.1111H}.

To follow the jet propagation, we consider a conical jet with a half-opening angle $\theta_j$ and a head cross section
\begin{equation}
    \Sigma_j(r_h)=\pi\theta_j^2r_h^2,
\end{equation}
and the relative strength of the jet and the surrounding fallback is characterized by the dimensionless luminosity
\begin{equation}
    \widetilde{L}(t,r_h)
    =
    \frac{L_j(t)}
    {\Sigma_j(r_h)\rho_{\rm pol}(r_h,t)c^3}.
\end{equation}
Notice that here the polar fallback is moving inward and its velocity in the merger frame can be estimated as 
\begin{equation}
    \beta_a(r,t)=-\frac{v_{\rm FB}(r,t)}{c},
\end{equation}
where $v_{\rm FB}>0$ is the free-fall velocity of the fallback material. Then the jet-head velocity in this moving medium can be approximated by \citep{2020MNRAS.491.3192H},
\begin{equation}
    \beta_h=
    \frac{1+\widetilde{L}^{-1/2}\beta_a}
    {1+\widetilde{L}^{-1/2}},
\end{equation}
and one could determine the trajectory of the jet head by $dr_h/dt=\beta_h c$. Here we initialize the secondary jet head at $r_h(t_{j,2})=R_d$, since the inner propagation time is negligible compared to the EE timescale.
Furthermore, we could define the crossing time $\Delta t_{\rm cross}$ as the time at which the jet head reaches the evolving outer edge of the fallback envelope,
\begin{equation}
    r_h(t_{j,2}+\Delta t_{\rm cross})
    =
    R_{\rm edge}(t_{j,2}+\Delta t_{\rm cross}).
\end{equation} 
The resulting elapsed time from jet launch to crossing $\Delta t_{\rm cross}$, head velocity $\beta_{h,\rm cross}$, and dimensionless luminosity $\widetilde{L}_{\rm cross}$ determine whether the secondary jet is choked or successfully emerges from the fallback envelope. Here we briefly summarize the adopted fiducial parameters and their values for the proposed fallback-fed secondary jet model in Table~\ref{table:parameters} and  determine the jet breakout regimes in Section~\ref{sec:regimes}.

\begin{table}
\caption{Parameters of the fallback-fed secondary jet model.}
\label{table:parameters}
\centering
\begin{tabular}{lcc}
\hline\hline
$M_c$
& Central-remnant mass
& $2.63\,M_\odot$ \\
$M_{\rm FB}$
& Total bound fallback mass
& $1.6\times10^{-3}\,M_\odot$ \\
$t_{\min}$
& Minimum fallback time
& $0.1\ {\rm s}$ \\
$\alpha$
& Fallback power-law index
& $5/3$ \\
$R_d$
& Reference radius
& $300M_\odot$ \\
$\eta_j$
& Accretion-to-jet efficiency
& $0.1$ \\
$t_{j,2}$
& Secondary-jet launch time
& $10\ {\rm s}$ \\
$t_{\rm eng}$
& Engine duration
& $100\ {\rm s}$ \\
\hline\hline
$f_{\rm circ}$
& Circularization fraction
& $[10^{-3},1]$ \\
$\theta_j$
& Jet half-opening angle
& $[0.01,0.50]\ {\rm rad}$  \\
$\xi_{\rm pol}$
& Polar refilling factor
& $1,\ 0.1,\ 0.01$ \\
\hline\hline
\end{tabular}
\end{table}

\begin{figure*}
\centering
\includegraphics[width=2.0\columnwidth]{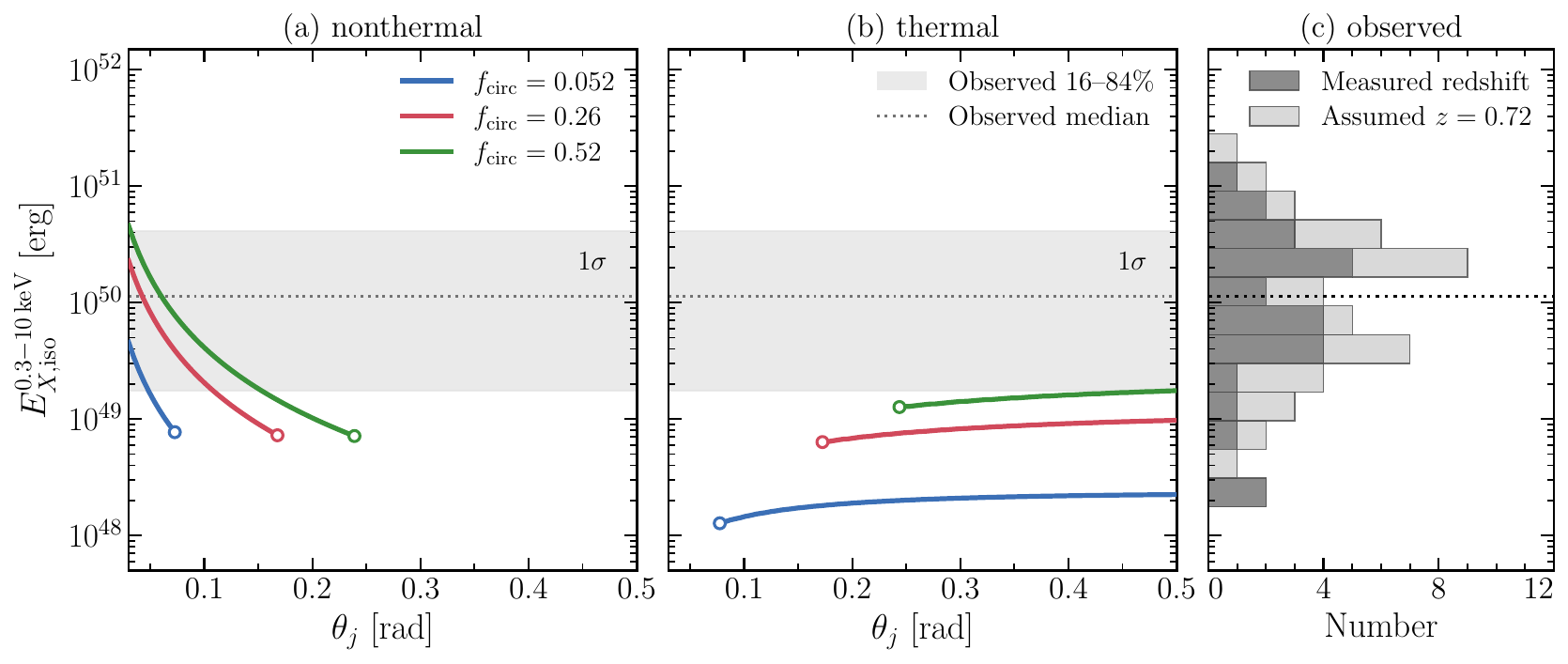}
\caption{Estimated and observed isotropic-equivalent 0.3--10 keV energies for the case $\xi_{\rm pol}=0.1$. Panels (a) and (b) show the jet/non-thermal dominated and cocoon-dominated/thermal dominated scenarios, respectively, as functions of $\theta_{j}$. Different colors denote the adopted circularization fractions obtained by combining the GRMHD-calibrated fraction of bound fallback not promptly captured, $f_{\rm npc}\sim 0.52$, with effective circularization efficiencies $\eta_{\rm circ}=0.1,0.5,1$. Each model curve is shown only where it satisfies the corresponding dynamical criterion and open endpoints mark the jet--cocoon-dominated boundary. The gray horizontal band and dotted line indicate the central 16-84\% interval and median of the observed extended-emission sample. Panel (c) shows the corresponding observed-energy histogram from \citet{2017ApJ...846..142K}. The dark and light histograms denote events with measured redshifts and events adopting $z=0.72$, respectively.}

\label{fig:Eiso}
\end{figure*}

\section{Jet breakout regimes and EE energetics}
\label{sec:regimes}

The fate of the secondary jet is determined by two conditions: (a) whether the relativistic jet material survives until the head reaches the outer edge, and (b) whether the jet-head remains dynamically strong at that time. Following the convention of \citet{2025ApJ...988...30H}, we quantify condition (a) by
\begin{equation}
    t_{\rm jet,cross}
    \equiv
    \frac{\Delta r_{\rm jet}}{c}
    =
    t_{\rm eng}
    -
    \Delta t_{\rm cross}(1-\bar{\beta_{h}}),
\end{equation}
where $\bar{\beta_{h}}$ is the time-averaged jet-head velocity and $\Delta r_{\rm jet}$ is the remaining distance between the jet-head and the jet-tail at the formal crossing time. The condition $t_{\rm jet,cross}>0$ indicates that a finite segment of relativistic jet material remains behind the head when it reaches the outer edge, whereas $t_{\rm jet,cross}\leq0$ implies that the tail catches the head before the crossing. As for condition (b), a simplified but natural approach is to estimate the dimensionless luminosity at the outer edge,
\begin{equation}
    \widetilde{L}_{\rm cross}
    \equiv
    \widetilde{L}
    \left[
    t_{j,2}+\Delta t_{\rm cross},
    R_{\rm edge}(t_{j,2}+\Delta t_{\rm cross})
    \right],
\end{equation}
which measures the relative strength of the jet energy flux to the rest-mass energy flux of the ambient fallback material. For $\widetilde{L}_{\rm cross}\ll1$, the head motion is controlled primarily by the inertia of the fallback material, and a large fraction of the jet power is transferred to the head and cocoon. For $\widetilde{L}_{\rm cross}\gg1$, the jet thrust dominates and the head propagates more rapidly with weaker coupling to the surrounding fallback. We therefore adopt $\tilde{L}_{\rm cross}=1$ as an approximate separator between fallback-dominated and jet-dominated heads. In summary, there are three possible outcomes:
\begin{itemize}
    \item $t_{\rm jet,cross}>0$ and
    $\widetilde{L}_{\rm cross}\geq1$:  jet-dominated  breakout;

    \item $t_{\rm jet,cross}>0$ and
    $\widetilde{L}_{\rm cross}<1$: cocoon-dominated breakout;

    \item $t_{\rm jet,cross}\leq0$: choked jet.
\end{itemize}
Therefore, our simplified toy model allows the relative importance of nonthermal jet-dominated emission and thermal cocoon-dominated emission to vary among events, and is qualitatively consistent with the spectral diversity observed in sGRB EE \citep[e.g.,][]{2024ApJS..275....9C}.

Figure~\ref{fig:phase_diagram} shows the dependency of the secondary jet breakout outcome on polar refilling factor $\xi_{\rm pol}$, half-opening angle $\theta_{j}$ and circularization fraction $f_{\rm circ}$. 
The overall trend follows the approximate scaling of dimensionless luminosity,
\begin{equation}
    \widetilde{L}
    \propto
    \frac{f_{\rm circ}}
    {\xi_{\rm pol}\theta_j^2}.
\end{equation}
At fixed $\xi_{\rm pol}$, on the one hand, increasing $f_{\rm circ}$ increases the circularized fallback
supply and hence the secondary-jet power also increases. On the other hand, decreasing $\theta_j$ concentrates the jet power into a smaller cross section and therefore favors a jet-dominated breakout. For a completely refilled polar region, $\xi_{\rm pol}=1$, the choked and cocoon-dominated regimes occupy a substantial fraction of the parameter space. Reducing the polar density to $\xi_{\rm pol}=0.1$ causes the choked region to contract and shifts the jet-dominated breakout boundary toward larger $\theta_{j}$ and smaller $f_{\rm circ}$. For $\xi_{\rm pol}=0.01$, the choked regime disappears within the explored parameter range, indicating that a strongly depleted polar cavity allows even relatively weak or wide jets to reach the outer edge. Moreover, the simulation-informed values of $f_{\rm circ}$
allow the regime boundaries to be translated into characteristic jet opening angles. For example, for the case of $f_{\rm circ}=0.26$, the jet to cocoon-dominated breakout transition occurs at
$\theta_j \simeq 0.17\,{\rm rad}$ when $\xi_{\rm pol}=0.1$. The critical angle decreases as the surviving polar fallback increases, approximately following the scaling $\theta_{j,\rm crit}\propto(f_{\rm circ}/\xi_{\rm pol})^{1/2}$.
Since the absolute value of $\xi_{\rm pol}$ cannot yet be determined self-consistently from the current simulations, we present 
continuous $\theta_{j,\rm crit}-\xi_{\rm pol}$ curves in Appendix~\ref{app:fcirc} for further reference.

Having determined the dynamical outcome of the secondary-jet breakout, we next provide an order-of-magnitude estimate of the observed X-ray energy in the  $0.3$--$10\ {\rm keV}$ band of the Swift X-ray telescope \citep[XRT,][]{2005SSRv..120..165B} for the jet-dominated and cocoon-dominated breakout cases. For a jet-dominated breakout, we assume that a typical fraction
$\eta_X=0.01$ of the surviving relativistic jet energy is converted into non-thermal X-rays. For a cocoon-dominated breakout, we estimate the thermal X-ray emission from the effective temperature of the jet cocoon. Details of these prescriptions are given in
Appendix~\ref{app:radiation}. Figure~\ref{fig:Eiso} shows the resulting isotropic-equivalent
X-ray energies for the fiducial case $\xi_{\rm pol}=0.1$ (the corresponding results for $\xi_{\rm pol}=0.01$ and $1.0$ are presented in Appendix~\ref{app:radiation}). Panels (a) and (b)
show the non-thermal and thermally-dominated scenarios, respectively, for different values of $f_{\rm circ}$, while panel (c) shows the observed EE energy distribution from \citet{2017ApJ...846..142K}. The non-thermal branch spans approximately
$E_{X,\rm iso}\sim10^{49}$--$10^{51}\ {\rm erg}$, whereas the
thermal branch lies mainly around
$E_{X,\rm iso}\sim10^{48}$--$10^{49}\ {\rm erg}$.
These energy scales broadly overlap those inferred for observed EE,
indicating that both successful-breakout branches are energetically
viable at the order-of-magnitude level. The thermal branch also yields
a characteristic temperature evolution in the soft-X-ray regime:
for a representative cocoon-dominated solution,
$k_{\rm B}T_{\rm eff}$ decreases from $\simeq17~{\rm keV}$ near
radiative breakout to $\simeq0.6~{\rm keV}$ by the end of the engine
activity, naturally passing through the $\sim1~{\rm keV}$ range.
This is comparable to the
$k_{\rm B}T_{\rm BB}\simeq0.8$--$1.1~{\rm keV}$ thermal component
reported for the EE of GRB~050724
\citep{2021MNRAS.501.4974V}. Finally, the choked-jet regime naturally
accommodates systems in which no secondary-jet EE emerges, as may be
relevant to sGRBs without detectable EE.

\section{Conclusions and discussion}
\label{sec:cons}
Motivated by GRMHD simulations showing that the late-time fallback in BNS mergers is approximately quasi-spherical, we propose a secondary-jet toy model in which fallback both fuels the jet through circularization and obstructs it through polar refilling. We find that a larger circularization fraction, a narrower jet, or a less-refilled polar channel favors a jet-dominated breakout, whereas
the opposite conditions lead to cocoon-dominated or choked outcomes. The estimated nonthermal and thermal X-ray energies broadly overlap the observed $0.3$--$10\ {\rm keV}$ energy scale of sGRB extended emission. This framework therefore provides a unified qualitative description of the observed diversity: jet-dominated and cocoon-dominated breakouts may favor predominantly nonthermal and thermal EE, respectively, while choked jets may produce no detectable EE. Overall, our results suggest that fallback-fed secondary jets are dynamically feasible and energetically compatible with the observed
EE phenomenon.

Note that we have made several assumptions and approximations in this model. In reality, there are many complexities that need to be considered for a more extensive investigation.  First, the association of jet-dominated and cocoon-dominated breakouts with non-thermal and thermal emission should be viewed as two limiting cases rather than  strict spectral predictions. Near the transition at $\widetilde{L}_{\rm cross}\sim1$, emission from both the surviving jet and the cocoon may contribute simultaneously. In addition, our estimates
of $E_{X,\rm iso}$ adopt simplified prescriptions of the X-ray efficiency and cocoon photosphere without explicitly modeling time-dependent radiative transfer
\citep[e.g.,][]{2016ApJ...831..175V,2026arXiv260309066X,2024Ap&SS.369..102T}. Despite these simplifications, our results nevertheless provide a first-order indication that the proposed model can reproduce the observed EE energy scale.

Second, the fallback properties adopted here are calibrated using a specific BNS GRMHD simulation and remain subject to associated uncertainties in the binary parameters and microphysics \citep{Chabanov:2024yqv}. In particular, the equation of state and the fate of the remnant may also affect the total fallback mass, angular-momentum distribution, and polar density structure. If a long-lived neutron star forms instead of a BH, its surface, magnetosphere, and spin-down power may substantially modify the fallback accretion flow and the conditions for
launching a secondary jet \citep[e.g.,][]{2018MNRAS.481.3670R,Musolino:2024sju,2025MNRAS.542..256B}.

Third, the primary goal of this work is to establish the dynamical and energetic feasibility of a secondary jet powered by quasi-spherical BNS fallback, rather than to reproduce individual EE light curves. Accordingly, we adopt a simplified prescription
for the secondary engine, treating $t_{j,2}$ phenomenologically and assuming that, once launched, the jet power tracks the circularized fallback supply,
$L_j\propto\dot{M}_{\rm FB}\propto t^{-5/3}$.
Our comparison with observations is therefore restricted to the integrated $0.3$--$10\ {\rm keV}$ energy scale and should not be
interpreted as a fit to the detailed temporal morphology. A self-consistent treatment of fallback circularization, disk evolution, jet-launching conditions, and time-dependent radiative
transfer is left for future work.

\section*{acknowledgement}
We thank Youjun Lu and Carlo Musolino for their suggestion and help in this work. ZC
acknowledges the Postdoctoral Fellowship Program of CPSF under Grant
Number GZB20250735.  MC and LR acknowledge the support from the ERC
Advanced Grant ``JETSET: Launching, propagation and emission of
relativistic jets from binary mergers and across mass scales'' (grant
no. 884631). LR also acknowledges the Walter Greiner Gesellschaft zur
F\"orderung der physikalischen Grundlagenforschung e.V.  through the Carl
W. Fueck Laureatus Chair.  \appendix
\section{Nuclear heating energy budget}\label{app:r-process}

\begin{figure}
    \centering
    \includegraphics[width=1.0\columnwidth]{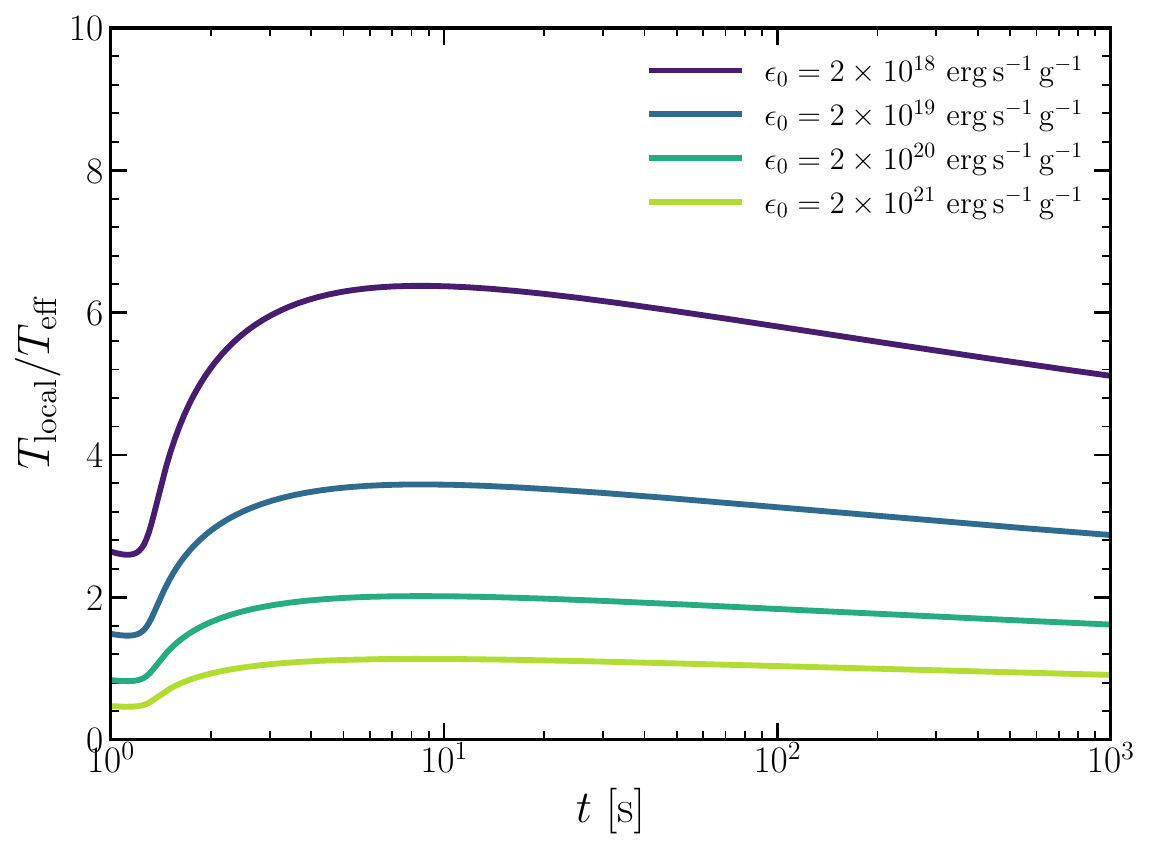}
    \caption{
  The ratio between the local gas temperature $T_{\rm local}(t)$ of the photosphere extracted from \texttt{Skynet} and the effective temperature $T_{\rm eff}(t)$. The purple, blue, teal and green lines show the results for different $r$-process heating-rate normalizations  $\epsilon_0=2\times 10^{18}, 2\times 10^{19},  2\times 10^{20}, 2\times 10^{21} \rm erg~s^{-1}~g^{-1}$, respectively. 
    }
    \label{fig:tlocalvseff}
\end{figure}

MDR24 proposed that the fallback material heated by radioactive $r$-process could directly reproduce the X-ray and gamma-ray emission observed during EE. In their radiation prescription, they adopt the local gas temperature of the nuclear reaction network \texttt{Skynet} \citep{2017ApJS..233...18L} to determine the emergent blackbody spectrum. However, a local thermodynamic temperature does not necessarily coincide with the effective temperature defined by the radiative flux. In this appendix, we examine whether the available radioactive-heating power can sustain the luminosity.

Following the standard treatment of radioactive heating in merger ejecta in the literature, 
\citep[e.g.,][]{2017ApJ...850L..37P, 2021MNRAS.505.1661B}, we first estimate the maximum radioactive-heating power available to the fallback material,
\begin{equation}
    L_{r}(t)
    =
    M_{\rm FB}(t)\,{\epsilon}_{\rm nuc}(t;\epsilon_0),
    \label{eq:radioactive_luminosity}
\end{equation}
where $M_{\rm FB}(t)$ is the fallback mass and ${\epsilon}_{\rm nuc}(t;\epsilon_0)$ is expressed as \citep{2012MNRAS.426.1940K}: 
\begin{equation}
\epsilon_{\rm nuc;\epsilon_0}(t)=\epsilon_{0}\epsilon_{Y_{\rm e}}(t)\left(\frac{\epsilon_{\mathrm{th}}(t)}{0.5}\right)\left[\frac{1}{2}-\frac{1}{\pi}\arctan\left(\frac{t-1.3\rm s}{0.11\rm s}\right)\right]^{1.3},
\end{equation}
where $t$ is measured in seconds, $\epsilon_0$ is the energy normalization and $\epsilon_{\rm th}$ is the thermalization efficiency given by \citet{2016ApJ...829..110B}:
\begin{equation}
\epsilon_{\mathrm{th}}(t)=0.36\left[\exp\left(\frac{-0.56t}{1\mathrm{day}}\right)+\frac{\ln\left(1+0.34(t/1\mathrm{day})^{0.74}\right)}{0.34(t/1\mathrm{day})^{0.74}}\right], 
\end{equation}
and $\epsilon_{Y_{\rm e}}\sim 1$ is a factor accounting for the dependence of radioactive heating on the electron fraction $Y_{\rm e}$ \citep{2015ApJ...813....2M}. Then, the effective radiative temperature allowed by this heating power is subsequently defined through,
\begin{equation}
    T_{\rm eff}(t)
    =
    \left[
    \frac{L_{r}(t)}
    {\sigma_{\rm SB}A_{\rm ph}(t)}
    \right]^{1/4},
\end{equation}
where  $\sigma_{\rm SB}$ is the Stefan--Boltzmann constant, $A_{\rm ph}(t)=4\pi R_{\rm ph}^2(t)$ is the emitting area and the photosphere radius $R_{\rm ph}(t)$ is kept the same as in MDR24.

Figure~\ref{fig:tlocalvseff} shows the ratio between  the local gas temperature $T_{\rm local}(t)$ of the photosphere extracted from \texttt{Skynet} and the effective temperature $T_{\rm eff}(t)$. As for the typical value of $\epsilon_0=2\times 10^{18} \rm erg~s^{-1}~g^{-1}$ suggested by r-process calculation from \citet{2012MNRAS.426.1940K}, the value of $T_{\rm local}$ obtained by MDR24 exceeds $T_{\rm eff}$ by a factor of $\gtrsim5$ over most of the EE-relevant timescales. Even with a higher $\epsilon_0\sim 2\times 10^{19} \rm erg~s^{-1}~g^{-1}$ inferred by AT2017gfo kilonova modeling \citep{2017ApJ...850L..37P}, this factor is about $\sim 4$. Therefore, interpreting $T_{\rm local}$ as the effective radiative temperature would imply a blackbody luminosity larger than the available radioactive-heating power by a factor $\sim (T_{\rm local}/T_{\rm eff})^4 \sim 10^2$--$10^3$. Moreover, photon diffusion through the optically thick fallback material can further suppress the emergent luminosity by a factor of $\sim 10$ \citep[e.g.,][]{1982ApJ...253..785A,2017ApJ...851L..21V,2020EPJA...56....8B}.  
Thus, under the adopted heating and diffusion assumptions, 
we estimate an EE luminosity of order $L_{\rm EE}\sim10^{43}$--$10^{44}\ {\rm erg\,s^{-1}}$,
below the luminosities typically inferred for EE in sGRBs
\cite{2017ApJ...846..142K}, suggesting that radioactive
heating alone is unlikely to sustain the observed emission.

\section{Fallback capture and circularization fractions}
\label{app:fcirc}

In this appendix, we briefly introduce how we estimate the fraction $f_{\rm npc}$ of fallback material that is not promptly captured and the circularization fraction $f_{\rm circ}$. We first use the \texttt{SFHO-q1.0} simulation of MDR24 to estimate $f_{\rm npc}$. The \texttt{SFHO-q1.0} model describes an equal-mass binary with $M_1=M_2=1.36\,M_\odot$ and employs the finite-temperature SFHO equation of state \citep{HEMPEL2010210}. The evolution was performed with the state-of-the-art \texttt{FIL} GRMHD code \citep{Rezzolla13}, the \texttt{Antelope} spacetime solver \citep{2019MNRAS.490.3588M}, and the \texttt{FIL-M1} neutrino-transport scheme \citep{2024MNRAS.528.5952M}. The remnant collapses to a BH $\sim 3.7\,{\rm ms}$ after merger.  Quasi-local measurements of the apparent horizon give $ M_{\rm BH}\simeq 2.63\,M_\odot$ and $\chi_{\rm BH}\equiv{J_{\rm BH}}/{M_{\rm BH}^2}\simeq0.78$ respectively. The BH spin is aligned with the positive $z$-axis, and we therefore set $\hat{\boldsymbol{s}}_{\rm BH}=\hat{\boldsymbol{z}}$ and $a=\chi_{\rm BH}M_{\rm BH}$. The corresponding outer horizon of the BH is located at $\bar r_+=1+\sqrt{1-\chi_{\rm BH}^2}$.

The properties of the outgoing material are recorded on a \texttt{HEALPix} \citep{2005ApJ...622..759G} detector at $R_d=300M_\odot$.  For detector interval $k$ and pixel $p$, the associated rest mass is
\begin{equation}
    \Delta m_{kp}
    =
    \left(\frac{d\dot M}{d\Omega}\right)_{kp}
    \Delta\Omega_p\,\Delta t_k ,
    \label{eq:detector_mass}
\end{equation}
where $\Delta\Omega_p$ is the solid angle represented by the pixel. To determine when each element returns, we follow MDR24 and estimate its specific orbital energy as $\epsilon_{kp}=-(1+u_{t,kp})$. An element is classified as bound when $\epsilon_{kp}<0$ and  its fallback time can be estimated by using the Newtonian Keplerian approximation
\begin{equation}
    t_{{\rm FB},kp}
    =
    \frac{\pi M_{\rm BH}}
    {\sqrt{2}\,|\epsilon_{kp}|^{3/2}},
    \label{eq:tfb_newtonian}
\end{equation}
which is found by MDR24 that only deviates from full Kerr-geodesic integrations by a factor less than $0.2\%$ for $t_{\rm FB}\gtrsim1\,{\rm s}$. 
Note here that we classify the ejecta into three orbital categories: Material with positive orbital energy is unbound and does not participate in the fallback. The gravitationally bound component is further divided into material on plunging orbits, which returns directly into the BH, and material that is not promptly captured, whose orbit possesses a turning point outside the horizon rather than plunging directly into the BH.
Only the latter provides a potential reservoir for subsequent circularization and accretion.

% \begin{figure}
%     \centering
%     \includegraphics[width=0.75\columnwidth]{figure5.pdf}
%     \caption{Orbital classification of the ejecta in the {GRMHD} BNS merger simulation. Approximately $50\%$ of the total ejecta is unbound (blue), while the remaining bound component is divided into plunging orbits
% ($\sim24\%$, orange) and material not promptly captured ($\sim26\%$). For illustration, the latter is further subdivided into
% circularized and non-circularized components with fractions $26\eta_{\rm circ}\%$ and $26(1-\eta_{\rm circ})\%$, respectively. \mc{Personal style-choice: The labels in the chart have different sizes, maybe increase the ones on the left and no worries it will fit when moving a bit more outside. }}
%     \label{fig:PIE}
% \end{figure}

To further estimate the value of $f_{\rm npc}$, we first construct the orbital constants in the remnant frame. Since our detector sphere is located at $R_d=300M_{\odot}$, we adopt the weak-field Newtonian approximation for tracking the dynamics of the fallback material. Here we note that though the detector is centered on the coordinate origin, the BH has a small coordinate displacement due to simulation uncertainties. Therefore, we define the position of a pixel relative to the black-hole centroid as
\begin{equation}
    \bar{\boldsymbol{x}}_{kp}
    =
    \boldsymbol{x}_p-\boldsymbol{x}_{\rm BH}(t_k),
    \label{eq:bh_centered_position}
\end{equation}
where $\boldsymbol{x}_{\rm BH}(t_k)$ is obtained by interpolating the apparent-horizon centroid to the detector time and  the overbar denotes a quantity evaluated relative to this BH frame. Similarly, in the weak-field detector region, we approximate the specific spatial momentum of the fluid element relative to the BH centroid as
\begin{equation}
    \bar{\boldsymbol{p}}_{kp}
    \simeq
    u^t_{kp}
    \left[
        \boldsymbol{v}_{kp}-\boldsymbol{v}_{\rm BH}(t_k)
    \right],
    \label{eq:bh_frame_velocity}
\end{equation}
where $\boldsymbol{v}_{\rm BH}=d\boldsymbol{x}_{\rm BH}/dt$ is the BH velocity. Then, the corresponding specific angular-momentum vector can be estimated as
\begin{equation}
    \boldsymbol{L}_{kp}
    =
    \bar{\boldsymbol{x}}_{kp}\times\bar{\boldsymbol{p}}_{kp},
    \qquad
    L_{kp}^2
    =
    \boldsymbol{L}_{kp}\cdot\boldsymbol{L}_{kp},
    \label{eq:angular_momentum_vector}
\end{equation}
and its component parallel to the BH spin is $L_{z,kp}
    = \boldsymbol{L}_{kp}\cdot\hat{\boldsymbol{s}}_{\rm BH}$.

At the detector radius, where $M_{\rm BH}/R_d\ll1$, we approximate the Kerr constants of motion per unit rest mass by
\begin{align}
    E_{kp} &\simeq -u_{t,kp}, \label{eq:energy_weakfield}\\
    Q_{kp} &\simeq
    L_{kp}^2-L_{z,kp}^2
    +a^2(1-E_{kp}^2)\cos^2\bar{\theta}_{kp},
    \label{eq:carter_weakfield}
\end{align}
where $\bar{\theta}_{kp}$ is the polar angle of the detector element measured from the black-hole spin axis,  $E$ is the conserved specific energy, $L_z$ is the conserved specific angular momentum, and $Q$ is the Carter constant.  The term $L^2-L_z^2$ is the weak-field limit of the angular momentum perpendicular to the spin axis, while the final term in Equation~(\ref{eq:carter_weakfield}) is the leading Kerr correction. Retaining $Q$ allows the prompt-capture criterion to account for inclined as well as equatorial trajectories.  We retain $E\simeq-u_t$, as in the fallback-time calculation of MDR24, while applying the remnant-frame correction to the spatial momentum used to construct $L_z$ and $Q$.

The approximate constants $(E,L_z,Q)$ obtained at the weak-field detector then specify a timelike Kerr geodesic per unit rest mass, whose radial turning points determine prompt capture. The dimensionless radial potential is
\begin{align}
    \mathcal R(\bar r)
    ={}&
    \left[
        E(\bar r^2+\chi_{\rm BH}^2)
        -\chi_{\rm BH}\ell_z
    \right]^2
    \nonumber\\
    &-
    \left(\bar r^2-2\bar r+\chi_{\rm BH}^2\right)
    \left[
        \bar r^2+(\ell_z-\chi_{\rm BH}E)^2+q
    \right],
    \label{eq:kerr_radial_potential}
\end{align}
where $\bar r=r_{\rm BL}/M_{\rm BH}$ is the dimensionless Boyer--Lindquist radial coordinate, $\bar r_d=R_d/M_{\rm BH}$, $\ell_z=L_z/M_{\rm BH}$, and $q=Q/M_{\rm BH}^2$, in geometrized units $G=c=1$.
For the selected element that is outward moving and bound, the radial potential must first satisfy an apocentre $\bar r_a>\bar r_d$.  After the element reverses its radial motion, we classify it as not promptly captured if the same allowed branch contains an inner turning point outside the horizon,
\begin{equation}
    \bar r_+<\bar r_p<\bar r_d,
    \qquad
    \mathcal R(\bar r_p)=0,
    \label{eq:external_pericentre}
\end{equation}
with $\mathcal R\geq0$ on the connected interval between $\bar r_p$ and $\bar r_a$. If the allowed branch instead extends continuously to the horizon, the element is classified as being on a plunging orbit. In summary, we encode this distinction through
\begin{equation}
    C_{{\rm npc},kp}
    =
    \begin{cases}
        1, & \text{not promptly captured},\\
        0, & \text{plunging orbit}.
    \end{cases}
    \label{eq:npc_indicator}
\end{equation}

Having assigned a Newtonian fallback time and a Kerr prompt-capture indicator to each valid detector element, we group the elements into fallback-time bins,
\begin{equation}
    \mathcal I_b
    =
    \left\{
        (k,p):
        t_{{\rm FB},kp}\in[t_{b,-},t_{b,+})
    \right\}, 
\end{equation}
and the total fallback rate and the fallback rate of material not promptly captured are defined as 
\begin{align}
    \dot M_{\rm FB}(t_b)
    &=
    \frac{1}{\Delta t_b}
    \sum_{(k,p)\in\mathcal I_b}\Delta m_{kp},
    \nonumber\\
    \dot M_{\rm npc}(t_b)
    &=
    \frac{1}{\Delta t_b}
    \sum_{(k,p)\in\mathcal I_b}
    \Delta m_{kp}C_{{\rm npc},kp}.
    \label{eq:fallback_rates}
\end{align}
Then, the corresponding fraction not promptly captured is
\begin{equation}
    f_{\rm npc}(t_b)
    =
    \frac{\dot M_{\rm npc}(t_b)}
    {\dot M_{\rm FB}(t_b)}.
    \label{eq:fnpc}
\end{equation}

\begin{figure}
    \centering
    \includegraphics[width=1.0\columnwidth]{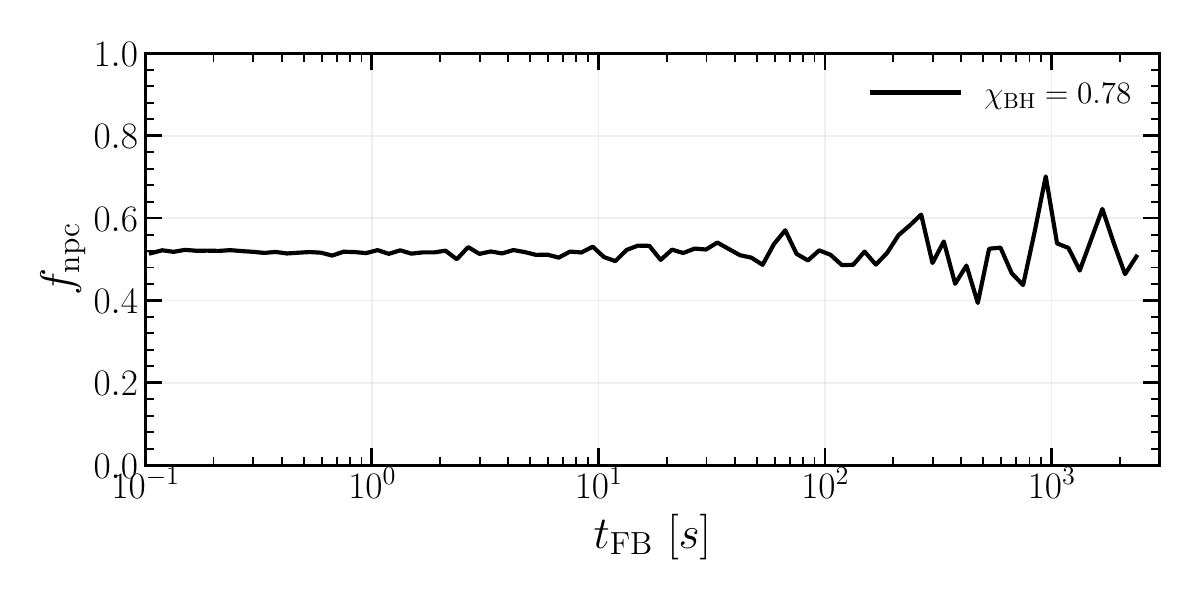}
    \caption{The time evolution of the fraction $f_{\rm npc}$ of fallback material not promptly captured, estimated from the \texttt{SFHO-q1.0} model reported in MDR24. 
    }
    \label{fig:npc}
\end{figure}

Figure~\ref{fig:npc} shows the estimated $f_{\rm npc}$ from the simulation data and it remains approximately constant up to $t_{\rm FB}\sim 10^3~\rm s$. 
Integrating over the time interval, the above classification gives  a time average value of, 
\begin{equation}
    f_{\rm npc}\simeq0.52,
    \label{eq:fnpc_result}
\end{equation}
where $f_{\rm npc}$ is defined relative to the bound fallback material. In terms of the total ejecta mass, approximately $\sim 50\%$
is unbound, while the bound component consists of $\sim24\%$ on plunging orbits
and $\sim26\%$ that is not promptly captured. Notably, we also find that the fraction of the fallback material not promptly captured that is located inside bipolar cones aligned with the black-hole spin axis is very small. For $\theta_j=0.1$, $0.2$, and $0.3~{\rm rad}$, the polar cones contain only $0.01\%$, $0.09\%$, and $0.6\%$ of the total fallback mass not promptly captured, respectively. Thus, even for the widest jet considered here, $\gtrsim99.3\%$ of this reservoir lies outside the polar cones. The reservoir of material not promptly captured is therefore overwhelmingly distributed outside the polar cones.

\begin{figure}
\centering
\includegraphics[width=1.0\columnwidth]{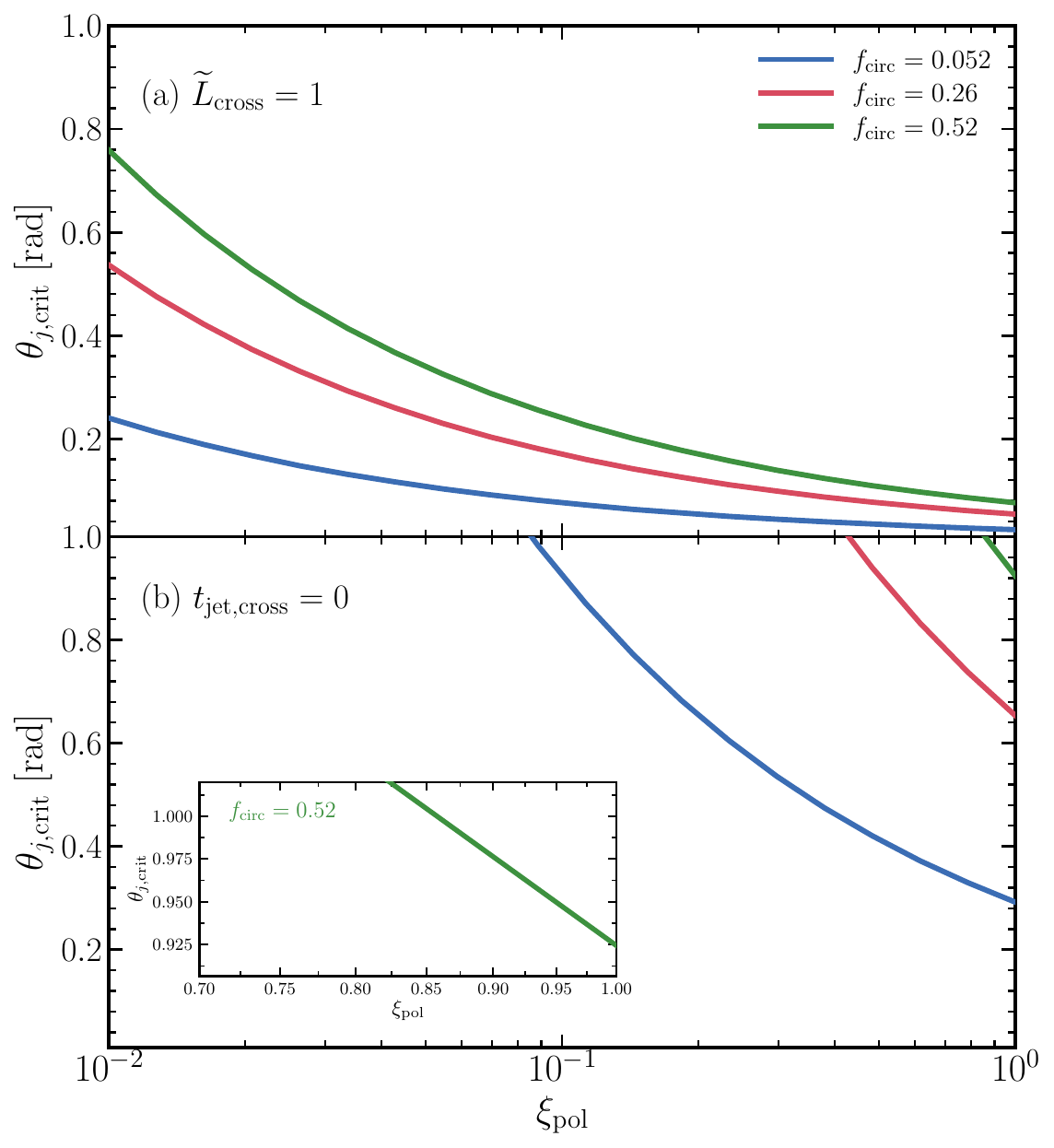}
\caption{
Critical jet half-opening angles as functions of the polar
refilling factor $\xi_{\rm pol}$ for $f_{\rm circ}=0.052$
(blue), $0.26$ (red), and $0.52$ (green).
The top and bottom panels show the jet--cocoon transition
($\widetilde{L}_{\rm cross}=1$) and the choking boundary
($t_{\rm jet,cross}=0$), respectively.
For each model, angles below the first boundary give
jet-dominated breakout, angles between the two boundaries
give cocoon-dominated breakout, and angles above the second
boundary give choked jets.
\label{fig:critical_angle}
}
\end{figure}

The component that is not promptly captured represents only the maximal reservoir available for circularization. Avoiding prompt capture is necessary but not sufficient for disk formation: the returning material must additionally dissipate orbital energy through stream interactions and shocks, redistribute angular momentum, and remain bound after dissipation. For illustration, we assume that only a fraction $\eta_{\rm circ}$ of this material is incorporated into the accretion flow. Since following these processes self-consistently would require
GRMHD simulations extending to much longer timescales than are currently computationally feasible, we write
\begin{equation}
    f_{\rm circ}
    =
    \eta_{\rm circ}f_{\rm npc},
    \qquad
    0\leq\eta_{\rm circ}\leq1,
    \label{eq:fcirc_parameterization}
\end{equation}
where $\eta_{\rm circ}$ is an effective circularization efficiency over the timescale relevant to the secondary engine.  An instructive, although not identical, analogue is provided by tidal disruption events (TDE) \citep[e.g.,][]{1975Natur.254..295H,1988Natur.333..523R,2021ARA&A..59...21G,2024ApJ...962....3C}: in both cases, weakly bound material returns on highly eccentric trajectories with an approximate rate of $\dot{M}_{\rm FB}\propto t^{-5/3}$, and its conversion into an accretion flow may depend on relativistic precession, stream--stream intersections, and shock dissipation.
TDE calculations nevertheless demonstrate that the outcome depends strongly on the stream geometry and on the time over which circularization is assessed.  Self-intersection shocks can circularize a coherent bound stream within approximately $1$--$10$ orbital periods \citep{2016MNRAS.455.2253B}.  In other configurations, the debris is incorporated more gradually into a spatially extended accretion flow \citep{2015ApJ...804...85S}, while fully relativistic calculations have found that most of the material can remain substantially eccentric for several fallback times and that formation of a compact disk may require several tens of fallback times \citep{2023ApJ...957...12R} . Although these TDE results cannot be mapped quantitatively onto the quasi-spherical BNS fallback studied here, they motivate considering both inefficient circularization and an optimistic limit in which most material not promptly captured joins the accretion flow. Hence, rather than assigning a preferred value that is not resolved by the simulation, we consider
\begin{equation}
    \eta_{\rm circ}=0.1,\ 0.5,\ 1,
\end{equation}
representing inefficient, intermediate, and maximal circularization. In this effective prescription, the three cases correspond to $10\%$, $50\%$, and $100\%$ of the material not promptly captured being incorporated into the accretion flow over the timescale relevant to the secondary engine. For $f_{\rm npc}=0.52$, these choices give $f_{\rm circ}\simeq0.052,\ 0.26,\ 0.52$ respectively. Figure~\ref{fig:critical_angle} shows the critical jet angles $\theta_{j,\rm crit}$ by fixing $f_{\rm circ}$ to these three values and varying $\xi_{\rm pol}$ continuously. As the polar channel becomes more strongly refilled, the jet encounters a denser fallback environment and a progressively narrower opening angle is required for a jet-dominated breakout. At $\xi_{\rm pol}=0.1$, the jet-dominated breakout boundary is approximately $\theta_j\simeq0.08$, $0.17$, and $0.24\ {\rm rad}$ for $f_{\rm circ}=0.052$, $0.26$, and $0.52$, respectively. The choked-jet boundaries lie above $\theta_j=0.3\ {\rm rad}$ for all three simulation-derived circularization fractions, showing that no choked jet is expected within our fiducial opening-angle range once $f_{\rm circ}\gtrsim0.08$.  For $f_{\rm circ}=0.26$ and $0.52$, the choking boundaries lie above the fiducial range $\theta_j\leq0.3\ {\rm rad}$. For $f_{\rm circ}=0.052$ and a fully refilled polar channel, the phase-diagram calculation instead admits choked jets near the upper end of this opening-angle range.

\section{X-ray Energy estimates}\label{app:radiation}

\begin{figure*}
\centering
\includegraphics[width=1.95\columnwidth]{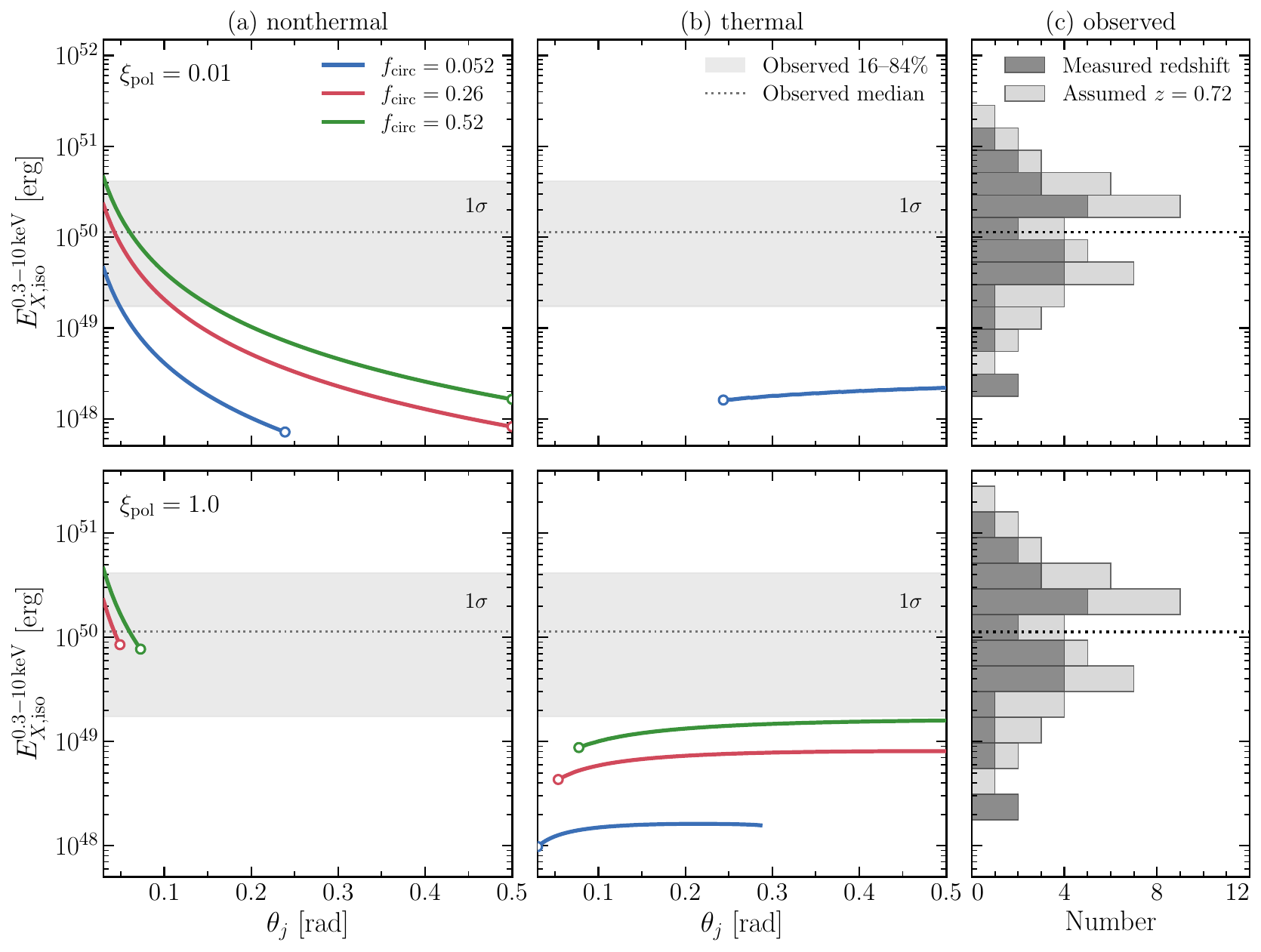}
\caption{The same as Figure~\ref{fig:Eiso}, but for
$\xi_{\rm pol}=0.01$ (top) and $\xi_{\rm pol}=1.0$ (bottom).}
 
\label{fig:c}
\end{figure*}

To connect the jet-breakout regimes with the observed energetics of EE, we roughly estimate the isotropic-equivalent energy in the $0.3$--$10\ {\rm keV}$ Swift-XRT band. At the crossing time, the amount of the engine time already consumed by the propagation of the jet-head is $t_{\rm lost}=\Delta t_{\rm cross}\left(1-\bar{\beta_{h}}\right)$. Then, the part of the jet that remains available to power escaping radiation was therefore injected between $t_{\rm inj,0}=t_{j,2}+t_{\rm lost}$ and $t_{\rm inj,1}=t_{j,2}+t_{\rm eng}$. 

For a jet-dominated breakout, we assume that a fraction $\eta_X=0.01$ of the surviving relativistic jet energy is converted into XRT emission \citep{2022ApJ...936L..10Z}.  Then, the corresponding isotropic-equivalent energy can be estimated as,
\begin{equation}
    E_{X,\rm iso}^{\rm NT}
    =
    \frac{\eta_X}{f_j}
    \int_{t_{\rm inj,0}}^{t_{\rm inj,1}}
    L_j(t_{\rm inj})\,dt_{\rm inj},
    \label{eq:nonthermal_energy}
\end{equation}
where $f_j=(1-\cos\theta_j)/2$ is the beaming fraction. 

For a cocoon-dominated breakout, we assume that the forward shock reaches the radiative surface, while the relativistic jet core remains terminated behind a persistently baryon-loaded head. The surviving engine power is transported outward predominantly as Poynting flux and continues to dissipate at the moving jet-head until the central engine shuts off. Note that we approximate the head velocity by the effective crossing value $\beta_{h,\rm cross}$ while neglecting the  detailed post-breakout acceleration or lateral expansion of the head.

The radiative-breakout time $t_{\rm bo}$ is defined as the first time at which
\begin{equation}
    \tau_{\rm ahead}
    \left[r_h(t_{\rm bo}),t_{\rm bo}\right]
    \equiv
    \kappa
    \int_{r_h(t_{\rm bo})}
         ^{R_{\rm edge}(t_{\rm bo})}
    \rho_{\rm pol}(r,t_{\rm bo})\,dr
    =
    \frac{1}{\beta_{h,\rm cross}},
    \label{eq:radiative_breakout_condition}
\end{equation}
where $\kappa=0.5\ {\rm cm^2\,g^{-1}}$ following MDR24. This criterion follows approximately from equating the photon-diffusion
time, $t_{\rm diff}\sim\tau_{\rm ahead}\Delta R/c$, with the jet-head
propagation time across the same layer,
$t_{\rm prop}\sim\Delta R/(\beta_{h,\rm cross}c)$. Thus, the corresponding radiative-breakout radius is $R_{\rm bo}\equiv r_h(t_{\rm bo})$.

After radiative breakout, we approximate the emitting head as a surface layer moving outward with a constant effective velocity $\beta_{h,\rm cross}c$. Its radius can therefore be estimated by,
\begin{equation}
    R_{\rm em}(t)
    =
    R_{\rm bo}
    +
    \beta_{h,\rm cross}c
    \left(t-t_{\rm bo}\right),
    \quad
    t_{\rm bo}\leq t
    \leq t_{j,2}+t_{\rm eng}.
    \label{eq:emitting_head_radius}
\end{equation}

The Poynting flux arriving at the slower moving head continuously dissipates a fraction of the jet power. Since the head itself moves outward at $\beta_{h,\rm cross}c$, the energy flux intercepted by the head is reduced by the relative factor
$(1-\beta_{h,\rm cross})$, and we approximate the deposited power as
\begin{equation}
    L_{\rm dep}(t)
    =
    \left(1-\beta_{h,\rm cross}\right)L_j(t).
    \label{eq:surface_deposited_power}
\end{equation}
Notably, the diffusion time across the remaining surface layer $t_{\rm diff,local}\sim{\tau_{\rm ahead}(R_{\rm edge}(t_{\rm bo})-R_{\rm bo})}/c$ is much shorter than the engine duration $t_{\rm eng}$. Thus, we could adopt the fast-cooling surface-heating limit for estimation, i.e.,  $L_{\rm bol}\simeq L_{\rm dep}$. Then the effective temperature can be estimated by,
\begin{equation}
    T_{\rm eff}(t)
    =
    \left[
    \frac{L_{\rm dep}(t)}
    {\sigma_{\rm SB}A_{\rm ph}(t)}
    \right]^{1/4},
    \label{eq:surface_effective_temperature}
\end{equation}
where $ A_{\rm ph}(t)=4\pi f_{j}R_{\rm em}^2(t)$ is the emitting area. For a blackbody of effective temperature $T_{\rm eff}$, the fraction of the bolometric luminosity falling in the XRT band is
\begin{equation}
    f_X(T_{\rm eff})
    =
    \frac{15}{\pi^4}
    \int_{0.3\,{\rm keV}/(k_{\rm B}T_{\rm eff})}
         ^{10\,{\rm keV}/(k_{\rm B}T_{\rm eff})}
    \frac{x^3}{e^x-1}\,dx.
    \label{eq:xrt_fraction}
\end{equation}
For an observer along the jet axis, the isotropic-equivalent
luminosity of the one-sided emitting spherical cap differs from its true luminosity by the geometrical factor $4A_{\rm proj}/A_{\rm ph}=2(1+\cos\theta_j)$,
where $A_{\rm proj}=\pi R_{\rm em}^2\sin^2\theta_j$ is the projected emitting area. The isotropic-equivalent thermal X-ray energy is therefore
\begin{equation}
    E_{X,\rm iso}^{\rm TH}
    =
    2(\cos{\theta_j}+1)
    \int_{t_{\rm bo}}^{t_{j,2}+t_{\rm eng}}
    L_{\rm dep}(t)
    f_X[T_{\rm eff}(t)]\,dt.
    \label{eq:thermal_energy}
\end{equation}

We estimate $E_{X,\rm iso}$ with Equations~(\ref{eq:nonthermal_energy}) and (\ref{eq:thermal_energy}) for the jet-dominated and cocoon-dominated breakout scenarios respectively. Although these simplified prescriptions do not include detailed particle acceleration, lateral cocoon expansion, or time-dependent radiation transfer, they retain the time dependence of the fallback-fed engine and
the thermal XRT band correction, providing a useful order-of-magnitude estimate of the expected X-ray energetics.

\bibliographystyle{apsrev4-2}
\bibliography{refs}% Produces the bibliography via BibTeX.

\end{document}